\documentclass[showpacs,showkeywords,aps,prl,preprint,superscriptaddress]{revtex4}

\usepackage{bm}
\usepackage{epsfig}

\begin{document}

\title{Experimental evidence of anapolar moments in the antiferromagnetic insulating phase of V$_2$O$_3$ obtained from x-ray resonant Bragg diffraction}

\author{J. Fern\'{a}ndez-Rodr\'{\i}guez}
\author{V. Scagnoli}
\author{C. Mazzoli}
\author{F. Fabrizi}

\affiliation{European Synchrotron Radiation Facility, BP 220, 38043  Grenoble Cedex, France}

\author{S.W. Lovesey}

\affiliation{ISIS Facility, Rutherford Appleton Laboratory, Oxfordshire OX11 0QX, United Kingdom} 

\affiliation{Diamond Light Source Ltd. Oxfordshire OX11 0QX, United Kingdom}

\author{J. A. Blanco}
\affiliation{Departamento de F\'{\i}sica, Universidad de Oviedo,
E-33007 Oviedo, Spain}

\author{D.S. Sivia}
\author{K.S. Knight}
\affiliation{ISIS Facility, Rutherford Appleton Laboratory, Oxfordshire OX11 0QX, United Kingdom}

\author{F. de Bergevin}
\author{L. Paolasini}
\affiliation{European Synchrotron Radiation Facility, BP 220, 38043  Grenoble Cedex, France}

\date{\today}

\begin{abstract}

We have investigated the antiferromagnetic insulating phase of the
Mott-Hubbard insulator V$_2$O$_3$ by resonant x-ray Bragg
diffraction at the vanadium K-edge. Combining the information obtained
from azimuthal angle scans, linear incoming polarization scans and  by
fitting collected data to the scattering amplitude derived from the
established chemical I2/a and magnetic space groups we provide evidence
of the ordering motif of anapolar moments (which results from parity
violation coupling to an electromagnetic field). Experimental data
(azimuthal dependence and polarization analysis) collected at
space-group forbidden Bragg reflections are successfully accounted
within our model in terms of vanadium magnetoelectric
multipoles. We demonstrate that resonant x-ray diffraction intensities
in all space-group forbidden Bragg reflections of the kind $(hkl)_m$
with odd $h$ are produced by an E1-E2 event. The 
determined tensorial parameters offer a test for ab-initio
calculations in this material, that can lead to a deeper and more
quantitative understanding of the physical properties of V$_2$O$_3$.

\end{abstract}



\maketitle

In the case of ions located in crystal positions that are not a center
of inversion symmetry, hybridization will occur between the valence
orbitals of that ion, thus enabling the possibility of observing
electronic transitions to the hybridized states via the mixed
dipole-quadrupole (E1-E2) channel in resonant x-ray
scattering~\cite{Marri2004,Lovesey2005}, which is sensitive to the
ordering of parity-breaking tensorial moments.  For example, the
ordering of magnetoelectric toroidal moments (anapoles) might be
observed.  An anapole moment characterizes a system that does not
transform into itself under space inversion.  These toroidal moments
were initially considered in the context of multipolar expansions in
nuclear physics~\cite{Dubovik90} and are related to a distribution of
magnetic fields which is quite different from those produced by
parity-even multipoles, such as dipole or quadrupole moments. The
magnetic field distribution of an anapole looks like the magnetic
field created by a current flowing in a toroidal winding, and the
field is completely confined inside the winding. Parity-breaking E1-E2
contributions~\cite{Templeton94} to scattering can be expressed in
terms of polar and magnetoelectric tensors that contain the anapole
operator~\cite{Lovesey2005}. The study of these
contributions to resonant x-ray diffraction is of fundamental
importance in current developments of the electronic structure of
materials with complex electronic properties, such as
magnetoelectricity, piezoelectricity and ferroelectricity that are of
potential technological interest ~\cite{Lottermoser2004}. The cross
correlation between magnetism and ferroelectricity in materials with
coexistence of spontaneous magnetization and polarization, termed
multiferroics, has recently become a subject of great scientific
impact.

Vanadium sesquioxide, V$_2$O$_3$, considered as a Mott-Hubbard
metal-insulator~\cite{Mott74} system, has been the object of intense
study from both
theoretical~\cite{Castellani78,Lovesey2002,Tanaka2002,Lovesey2007a}
and experimental~\cite{Paolasini99, Paolasini2001,Park2000,Hague2008}
points of view in the last decades. This compound has an interesting
phase diagram, with an antiferromagnetic insulator phase (AFI) at low
temperatures, and a paramagnetic metallic one (PM) above the N\'{e}el
temperature ($T_N \approx 150$~K). The metal-insulator transition is
accompanied by a strongly first-order structural phase transition in
which the room-temperature corundum structure (R$\bar{3}$c) is
modified to monoclinic I2/a. In 1978 Castellani, Natoli and
Ranniger~\cite{Castellani78} proposed a theoretical model to explain
the magnetic structure in the AFI phase from the ordering pattern of
the occupation of the $t_{2g}$ (degenerate) orbitals. This model was
considered valid until 1999, when resonant x-ray diffraction
(Paolasini et al.~\cite{Paolasini99}), and magnetic dichroism
measurements (Park et
al.~\cite{Park2000}) demonstrated that the spin of the Vanadium atoms
was $S_V=1$, whereas the model of Castellani predicted $S_V=1/2$.  

The magnetic structure is such that the I-centering cell translation
is time inverting.  Because of the magnetic moments being colinear in
the $\mathbf{a}_m$-$\mathbf{c}_m$ plane, normal magnetic peaks are
observed only at reflections $(hkl)_m$ with even $h$ even and odd $(k+l)$.
Yet, peaks have been measured at odd $h$, even $(h+l)$, at the
resonant prepeak of Vanadium K-edge.  They were initially interpreted
as produced by orbital (time-even)
ordering~\cite{Paolasini99,Paolasini2001}. However, analyses of the
observations carried out by Lovesey {\it et al.}~\cite{Lovesey2002}
and Tanaka~\cite{Tanaka2002} demonstrated that the resonant Bragg
diffraction intensities are produced by the ordering of magnetic
(time-odd) V multipoles.  There has been controversy on the dominant
resonant event producing the measured intensisties, having been
proposed a parity-even E2--E2 event~\cite{Lovesey2002}, a parity-odd
E1--E2 event~\cite{Tanaka2002}, or a combination of
them~\cite{Joly2004,Lovesey2007a}.

In this paper we present measurements at space-group forbidden Bragg
reflections $(hkl)_m$ with odd $h$ in the low temperature monoclinic
phase of V$_2$O$_3$.  Together with energy profiles and azimuthal
dependence, we measured the polarization dependence at a fixed
azimuthal point varying the incident linear polarization with a
phase-plate setup. In this way, we can obtain the Stokes parameters
for the secondary beam as a function of both incident and scattered
linear polarization angle, without moving the crystal. This method can
elucidate the presence of resonances that are very close in energy,
playing on their relative phase shifts~\cite{Mazzoli2007}. Experiments
were carried out at the ID20 beamline~\cite{Paolasini2007} of the
ESRF. A single crystal of 2.8\% Cr-doped (V$_{1-x}$Cr$_x$)$_2$O$_3$
($x = 0.028$) was mounted on the four-circle diffractometer with
vertical scattering geometry. For the varying linear incoming
polarization, a diamond phase-plate of thickness 300~$\mu$m was
inserted into the incident beam. The phase plate was operated in
half-wave plate mode to rotate the incident polarization into an
arbitrary plane~\cite{Giles95,Scagnoli2009} described by Stokes
parameters $P_1 = cos(2\eta )$ and $P_2 = sin(2\eta )$, being $\eta$
the angle between the incident beam electric field and the axis
perpendicular to the scattering plane, i.e., $\eta=0$ corresponds to
polarization perpendicular to the scattering plane ($\sigma$
polarization). The Poincar\'{e}-Stokes parameters are defined by $P_1
= (|\varepsilon_{\sigma}|^2 - |\varepsilon_{\pi}|^2)/P_0 , P_2 = 2
\mathrm{Re} (\varepsilon_{\sigma}^*\varepsilon_{\pi})/P_0, P_3 = 2
\mathrm{Im} (\varepsilon_{\sigma}^*\varepsilon_{\pi})/P_0$, with $P_0
= |\varepsilon_{\sigma}|^2 + |\varepsilon_{\pi}|^2$ the total
intensity, and where $\varepsilon_{\sigma}$ and $\varepsilon_{\pi}$
are the components of the beam polarization vector perpendicular and
parallel to the scattering plane.

Fig. 1 shows the energy profiles for the $(30\bar{2})_m$ and
$(10\bar{2})_m$ reflections. They contain a single resonant peak
centered around 5.465 keV, as it is expected for reflections $(hkl)_m$
with odd $h$, in which the resonant peak from 5.47 to 5.49 keV, ascribed
to an $E1-E1$ event is
forbidden~\cite{Lovesey2002,Paolasini99}. Energy profiles in Fig. 1
show the presence of a shoulder at E=5.4665 keV for both reflections,
which opens the possibility for the existence of two lorentzians
separated by approximately 2 eV contributing to the observed
intensity. This energy separation can be related to crystalline
electric field energy transfer. The value of 10Dq was estimated as 2.1
eV from RIXS measurements~\cite{Hague2008}.  In order to
elucidate the possibility of interference of different lorentzians we
have performed polarization analysis measurements of the dependence of
the Stokes parameters of the secondary beam with the angle of primary
linear polarization $\eta$ · using the phase plate (Fig. 2).  In the
case of the $(10\bar{2})_m$ reflection we have repeated the
polarization analysis measurements at 3 different energies of the
incoming x-rays (Fig. 2-b). The presence of multiple lorentzians at
different energies with different tensorial properties would be
revealed by the appearance of circular polarization in the
polarization measurements, which can be estimated from the dependence
with linear incident polarization of the measured secondary Stokes
parameters $P'_1$ and $P'_3$, and by a variation in the shape of the
Stokes parameter curves when the energy of the x-rays is changed. We
try a fit of the polarization data with a real Jones matrix derived
from the unit-cell structure factors for the different polarizations
$F_{\sigma-\sigma'}$, $F_{\pi-\sigma'}$, $F_{\sigma-\pi'}$,
$F_{\pi-\pi'}$ (formulae for the outgoing Stokes parameters can be
seen in ref.~\onlinecite{Fernandez2008}) together with a
depolarization factor $(1-1/2 d \sin^2 \eta$) that multiplies the
linear components $N_1$ and $N_2$ of the Stokes vector of the incoming
light, being $d$ an adjustable parameter.  This depolarization
produced by the phase plate depends on the angle $\eta$ indicating the
angle of linear polarization of the incoming beam and its maximum is
expected at $\eta=90$ degrees. From the fitting to the $(30\bar{2})_m$
data we obtain the following parameters,
\begin{eqnarray}
d &=& 0.162 \pm 0.011 \nonumber \\
F_{\pi-\pi'}/F_{\sigma-\sigma'} &=& 0.8  \pm  0.3 \nonumber \\
F_{\pi-\sigma'}/F_{\sigma-\sigma'} &=& 0.64 \pm 0.02 \nonumber \\
F_{\sigma-\pi'}/F_{\sigma-\sigma'} &=& -3.01  \pm  0.05,   
\end{eqnarray}
and from the fitting to the $(10\bar{2})_m$ polarization data we obtain,
\begin{eqnarray}
d &=& 0.362 \pm 0.014 \nonumber \\
F_{\pi-\pi'}/F_{\sigma-\sigma'} &=& 0.099 \pm 0.005 \nonumber \\
F_{\pi-\sigma'}/F_{\sigma-\sigma'} &=& 1.20 \pm 0.01 \nonumber \\
F_{\sigma-\pi'}/F_{\sigma-\sigma'} &=& 0.770 \pm 0.005.
\end{eqnarray}
In both cases the magnitudes of the depolarization factor $d$ is
similar to that estimated from preliminary measurements with the phase
plate.  We obtain two different values of the depolarization d due to
the fact that polarization measurements in the $(30\bar{2})_m$ and in
the $(10\bar{2})_m$ reflections were performed in different
experiments. The good agreement obtained in the fittings for both
reflections with experimental data for a model in which we take into
account the depolarization introduced by the phase plate, together
with the fact that there is no significative dependence of the shape
of Stokes parameters curves when the energy of the x-rays is changed
in the measurements of $(10\bar{2})_m$ reflection leads us to conclude
that there is no evidence of the appearance of circular polarization
and that the measured intensities in both reflections are produced by
single oscillators with the same tensorial character. All of this
supports the validity of fitting the data with a single oscillator
model~\cite{Lovesey2007a}.  In this aspect, the model that will be used
to describe experimental data differs to the case of K$_2$CrO$_4$,
where there is a strong evidence of the appearance of circular
polarization~\cite{Mazzoli2007,Fernandez2008} and the intensities
were consequently modelled in terms of different lorentzians centered
at different energies for the different resonant events.

In Fig. 3 we show the azimuthal dependence of $(30\bar{2})_m$ and
$(10\bar{2})_m$ reflections measured at T=100 K in the AFI phase.  In order
to eliminate the effect of the absorption in the azimuthal curves,
data have been corrected according to the formula
\begin{eqnarray}
I_{\mathrm{corr}}&=& I_{\mathrm{obs}}(1+\sin \alpha_0 /\sin\alpha_1), \nonumber
\end{eqnarray}
where $\alpha_0$ and $\alpha_1$ are the incident and reflected angles
of the beam with the sample. The measured azimuthal dependence of
$(30\bar{2})_m$ shows a satisfactory agreement with previously
published data~\cite{Lovesey2007a}, as it can be seen in Fig. 3-a.

The azimuthal dependence and polarization data collected for
$(10\bar{2})_m$reflection present a good agreement with the expression for
the parity-breaking event resonant x-ray scattering $F(E1-E2)$ presented
in ref.~\onlinecite{Lovesey2007a}, being possible to fit together polarization scans
curves and azimuthal data.  The azimuthal and polarization data
for the $(30\bar{2})_m$ reflection can also be fitted to the $E1-E2$
structure factor expressions together with $(10\bar{2})_m$ data in terms of
the same set of tensorial parameters, which permits us to conclude that
intensities in both $(30\bar{2})_m$ and $(10\bar{2})_m$ are produced by an
$E1-E2$ event. The result of the fitting is shown in Fig. 3. The
determined parameters from fitting $|F(E1-E2)|^2$  to the data are,
\begin{eqnarray}
\mathrm{Im}\langle G ^1_1\rangle  /\mathrm{Im}\langle  G^3_3\rangle&=&  0.263 \pm 0.013  \nonumber \\
\langle G^1_0 \rangle /\mathrm{Im}\langle  G^3_3\rangle&=&          3.48     \pm 0.06  \nonumber \\
\mathrm{Im}\langle G^2_2 \rangle /\mathrm{Im}\langle  G^3_3\rangle &=& -8.38  \pm 0.02  \nonumber \\
\mathrm{Re}\langle G^2_1\rangle  /\mathrm{Im}\langle  G^3_3\rangle &=& 3.908  \pm 0.013  \nonumber \\
\mathrm{Re}\langle G^3_2 \rangle /\mathrm{Im}\langle  G^3_3\rangle  &=& -0.205	 \pm 0.007 \nonumber \\
\mathrm{Im}\langle G^3_1 \rangle /\mathrm{Im}\langle  G^3_3\rangle &=& -3.410	 \pm 0.015 \nonumber \\
\langle G^3_0\rangle /\mathrm{Im}\langle  G^3_3\rangle &=&  -4.33 \pm 0.03. \nonumber \\
\end{eqnarray}			      	  
Values for $\mathrm{Im}\langle G ^1_1\rangle$ and $\langle G^1_0
\rangle$ are direct estimates of the orbital anapolar moment $\langle
\mathbf{\Omega}\rangle$, $\mathrm{Im}\langle G^3_3\rangle$, $\mathrm{Re}\langle
G^3_2 \rangle$, $\mathrm{Im}\langle G^3_1 \rangle$ and $\langle
G^3_0\rangle$ are estimates of the moment $\langle
{(\mathbf{L}\otimes (\mathbf{L}\otimes\mathbf{\Omega})^2})^3\rangle$, 
where $\mathbf{L}$ is the operator for orbital
angular momentum, and $\mathrm{Im}\langle G^2_2 \rangle$ and
$\mathrm{Re}\langle G^2_1\rangle$ are estimates of the moment $\langle
(\mathbf{L} \otimes \mathbf{n})^2\rangle$, where $\mathbf{n}$
is the polar unit vector.  Operators $\mathbf{\Omega}$ and
$({\mathbf{L}\otimes(\mathbf{L}\otimes\mathbf{\Omega})^2})^3$ are true
spherical tensors while $(\mathbf{L} \otimes \mathbf{n})^2$ is a
pseudo-spherical-tensor. Additional information about parity-odd
tensors may be found in
refs.~\onlinecite{Lovesey2005,Dubovik90,Carra2000,Carra2003}. The presence
of a parity-even contribution $F(E2-E2)$ to resonant x-ray scattering in
this kind of reflections has been
suggested~\cite{Lovesey2002,Joly2004,Lovesey2007a} and within measured
reflections its strongest contribution would appear in the reflection
$(30\bar{2})_m$ as $F(E2-E2)$ is weighted by a global multiplicative
factor $\sin \nu$, with $\nu =hx + ky + lz$, being $x$, $y$, $z$
crystallographic parameters and the $h$, $k$, and $l$ the Miller
indices of the reflection~\cite{Lovesey2007a}. However, our fitting
describes data in $(30\bar{2})_m$ with great precision, which makes us
conclude that parity-even $E2-E2$ contribution is absent to a good
degree of approximation. From the fitted parameters, we can determine
the direction of the projection in the $\mathbf{a}_m$-$\mathbf{c}_m$
monoclinic plane of the anapolar moments of the Vanadium ions, which
from the quotient between $\mathrm{Im}\langle G ^1_1\rangle$ and
$\langle G^1_0 \rangle$ is estimated as forming 4 degrees with the
$(10\bar{1})_m$ reciprocal lattice vector. In Fig. 4, we show the
projection in the $\mathbf{a}_m$-$\mathbf{c}_m$ plane of the anapolar moments
of the eight vanadium ions in the monoclinic unit cell, together with
the arrangement of magnetic moments~\cite{Moon70} established below
$T_N$.

In conclusion, our results demonstrate that resonant x-ray diffraction
at the V K-edge for space-group forbidden Bragg reflections of the
kind $(hkl)_m$ with odd $h$ is produced by a parity-breaking E1-E2
event, being the contribution from parity-even transitions absent to a
very good degree of approximation. Polarization analysis measurements
at the $(30\bar{2})_m$ and $(10\bar{2})_m$ reflections probe the fact
that the intensities measured are coming with a good degree of
approximation from single lorentzians. Experimental data (azimuthal
variation and polarization analysis) collected at different
space-group forbidden Bragg reflections are successfully accounted
within our model in terms of values for the V magnetoelectric
multipoles.  The derived tensorial parameters include direct estimates of
expectation values of the vanadium anapolar moment and other
magnetoelectric moments~\cite{Lovesey2007a}. These results solve the
controversy on the origin of the resonant x-ray diffraction intensities
in V$_2$O$_3$~\cite{Lovesey2002,Tanaka2002,Joly2004,Lovesey2007a}. The
derived tensorial parameters offer a test for
ab-initio calculations of the electronic structure of vanadium
sesquioxide that can lead to a deeper and more quantitative
understanding of its electronic properties.

We acknowledge useful discussions with Carsten Deflets. Financial
support has been received from Spanish MECD Grant No.
MAT2008-06542-104-03. One of us, J.F.R. is grateful to Gobierno del
Principado de Asturias for the financial support of a Postdoctoral
grant from Plan de Ciencia, Tecnolog\'{\i}a e Innovaci\'{o}n (PCTI) de Asturias
2006-2009.

\begin{figure}
\begin{center}
\includegraphics[width=3in]{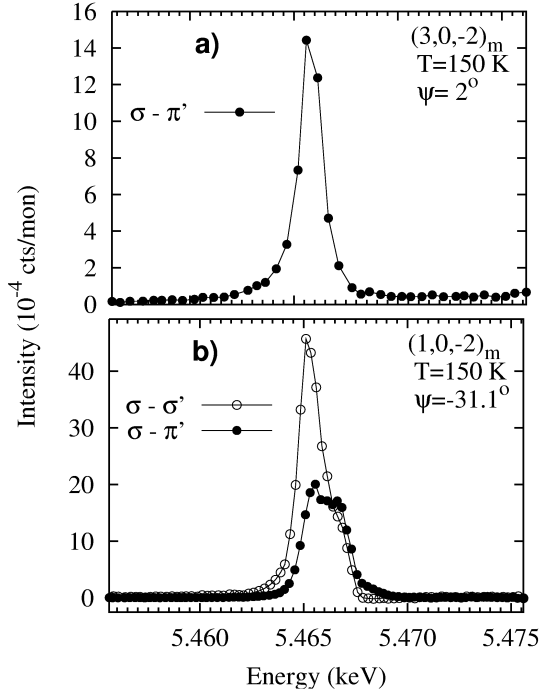}
\end{center}
\caption{\label{Fig1} Measured energy profiles of the $(30\bar{2})_m$
reflection (a) at the azimuthal angle of $\psi = 2$ degrees and
$(10\bar{2})_m$ reflection (b) at $\psi = -31.1$ degrees. The origin
of the azimuthal angles corresponds to $(010)_m$ reflection in the
plane of scattering.  }
\end{figure}

\begin{figure}
\begin{center}
\includegraphics[width=6in]{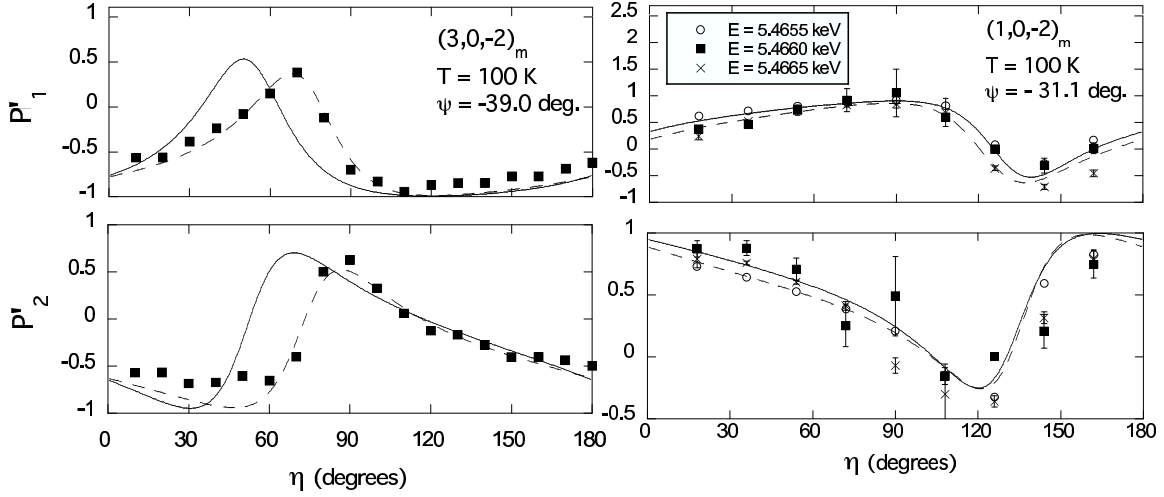}
\end{center}
\caption{\label{Fig2} Linear incident polarization dependence of the
Stokes parameters $P'_1$ and $P'_2$ of the secondary beam  in the
reflections $(30\bar{2})_m$ ($\psi = -39.0$ degrees) collected at T=100 K with
an incident energy of the x-rays $E = 5.4660$~keV and for the reflection
$(10\bar{2})_m$ at $\psi = -31.1$ degrees and T=150 K and at three
different energies $E=5.4655$, $5.4660$ and $5.4665$ keV. Dashed line
corresponds to the fitting of the data to a general Jones matrix
produced by a single oscillator including the effect of the phase
plate depolarization. Continuous line corresponds to the fitting of
polarization data together with azimuthal scans using the tensorial
parameters presented in eq. (3). In the origin of the azimuthal angle
the $(010)_m$ reflection is in the plane of scattering.  }
\end{figure}

\begin{figure}
\begin{center}
\includegraphics[width=6in]{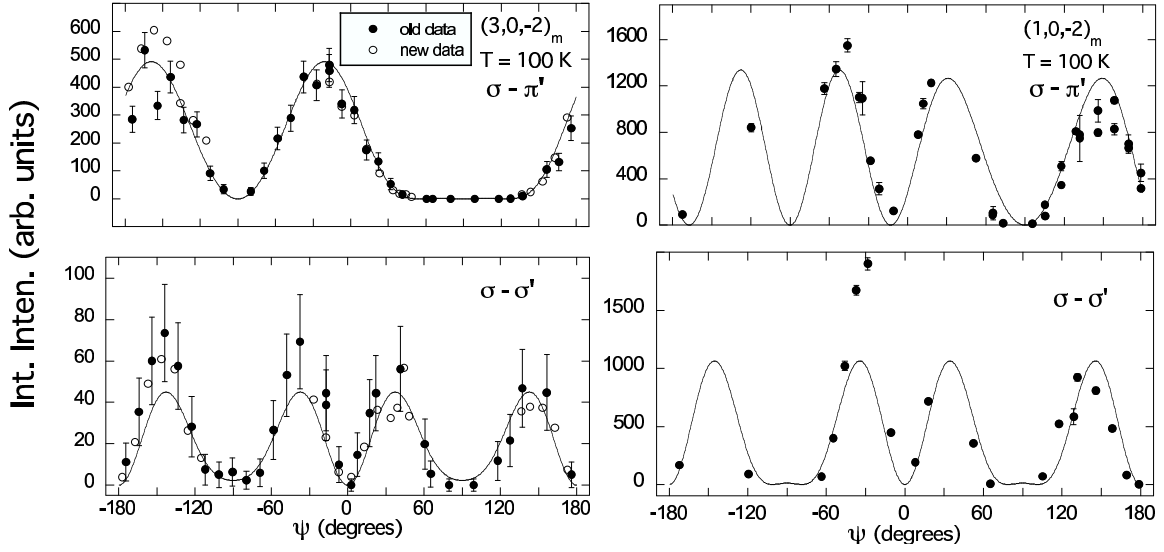}
\end{center}
\caption{\label{Fig3} Azimuthal angle scans measured on the
reflections $(30\bar{2})_m$ and $(10\bar{2})_m$ at $T=100$ K with
energy $E = 5.46627$ keV.  In the case of $(30\bar{2})_m$ we show the
agreement of new data with previously published
data~\cite{Lovesey2007a}. The continuous line correspond to the fitting
to the expressions for the $F(E1-E2)$ scattering length presented
in~\cite{Lovesey2007a} with the tensorial parameters shown in
eq. (3). At the
origin of the azimuthal angle $(010)_m$ reflection is in the plane
of scattering. Both polarization channels $\sigma-\sigma'$ and
$\sigma-\sigma'$ are shown.  }
\end{figure}

\begin{figure}
\begin{center}
\includegraphics[width=3.5in]{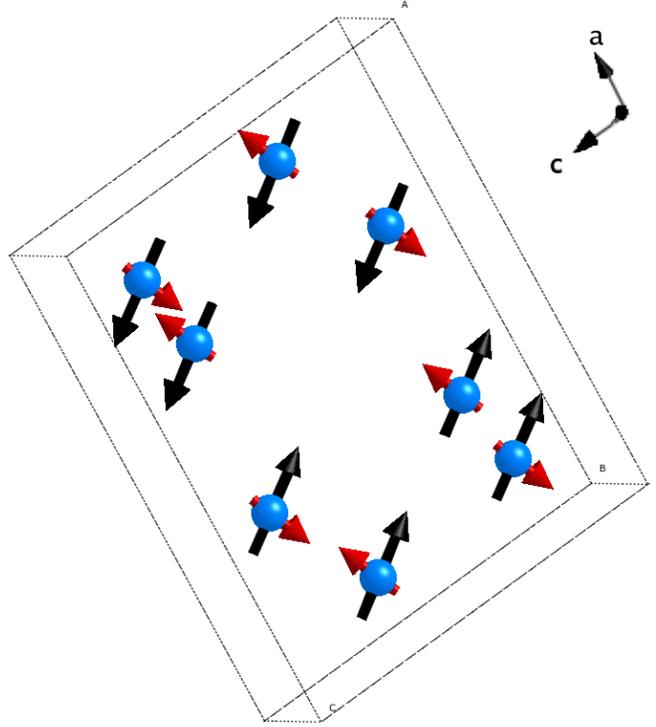}
\end{center}
\caption{\label{Fig4}
(color online) Positions of the Vanadium ions in the monoclinic unit cell
adopted by V$_2$O$_3$ below the Neel temperature,  together with the
configuration of the magnetic moments~\cite{Moon70} (red arrows)
and the determined projection of their anapolar moments (black arrows)
in the $\mathbf{a}_m$-$\mathbf{c}_m$ plane. The basis 
vector $\mathbf{b}_m$ is normal to the plane of the diagram. 
}
\end{figure}

\bibliography{V2O3_2009}

\end{document}